\documentclass[prd,showpacs,floatfix,twocolumn,,amsmath,amssymb,floatfix]{revtex4}
\usepackage{graphicx,color,dcolumn,booktabs,bm}
\usepackage{longtable,lscape}
\usepackage{txfonts}
\usepackage{overpic}
\usepackage{amssymb}
\usepackage{indentfirst}
\usepackage{epsfig}
\usepackage{feynmf}   
\usepackage{epstopdf}   
\usepackage{slashed}  
\usepackage{cases}
\usepackage{color}
\usepackage{multirow}
\usepackage{graphicx,color,dcolumn,booktabs,bm}

\graphicspath{{Figures/}} %

\graphicspath{{Figures/}} %
\usepackage[colorlinks, citecolor=blue,anchorcolor=red,menucolor=red, linkcolor=red,filecolor=red,runcolor=red,urlcolor=blue,frenchlinks=red]{hyperref}

\begin{document}

\title{The $D$-wave heavy-light mesons from QCD sum rules}

\author{Dan Zhou$^1$}
\author{Er-Liang Cui$^1$}
\author{Hua-Xing Chen$^1$\footnote{Corresponding author}}
\email{hxchen@buaa.edu.cn}
\author{Li-Sheng Geng$^1$}
\email{lisheng.geng@buaa.edu.cn}
\author{Xiang Liu$^{2,3}$\footnote{Corresponding author}}
\email{xiangliu@lzu.edu.cn}
\author{Shi-Lin Zhu$^{4,5,6}$\footnote{Corresponding author}}
\email{zhusl@pku.edu.cn} \affiliation{
$^1$School of Physics and Nuclear Energy Engineering and International Research Center for Nuclei and Particles in the Cosmos, Beihang University, Beijing 100191, China \\
$^2$School of Physical Science and Technology, Lanzhou University, Lanzhou 730000, China\\
$^3$Research Center for Hadron and CSR Physics, Lanzhou University and Institute of Modern Physics of CAS, Lanzhou 730000, China\\
$^4$School of Physics and State Key Laboratory of Nuclear Physics and Technology, Peking University, Beijing 100871, China \\
$^5$Collaborative Innovation Center of Quantum Matter, Beijing 100871, China \\
$^6$Center of High Energy Physics, Peking University, Beijing
100871, China }

\begin{abstract}
We study the $D$-wave $\bar c s$ heavy meson doublets $(1^-,2^-)$
and $(2^-,3^-)$ using the method of QCD sum rule in the framework of
heavy quark effective theory. Choosing the same threshold values
$\omega_c$ around 2.7 Gev, we calculate the masses of the $1^-$ and
$3^-$ states. They are $m_{D_{s1}^*}$ = $2.81 \pm 0.10$ GeV and
$m_{D_{s3}^*}$ = $2.85 \pm 0.08$ GeV, consistent with the newly
observed $D_{s1}^*(2860)$ and $D_{s3}^*(2860)$ states by LHCb. The
masses of their $2^-$ partners are calculated to be $2.82 \pm 0.10$
and $2.81 \pm 0.08$ GeV. The mass splittings within the same doublet are calculated to be $m_{D_{s2}}
- m_{D_{s1}^*} = 0.016 \pm 0.007$ GeV and $m_{D_{s3}^*} -
m_{D^\prime_{s2}} = 0.039 \pm 0.014$ GeV.
\end{abstract}

\maketitle

\section{Introduction}

Since the observation of $D_{s0}^*(2317)$ in 2003
\cite{Aubert:2003fg}, more and more charmed-strange mesons have been
reported experimentally, which include $D_{s1}(2460)$
\cite{Besson:2003cp}, $D_{s1}(2710)$
\cite{Brodzicka:2007aa,Aubert:2009ah}, $D_{sJ}(2860)$
\cite{Aubert:2006mh,Aubert:2009ah}, and $D_{sJ}(3040)$
\cite{Aubert:2009ah} (see Ref. \cite{Liu:2010zb} for a concise
review). Very recently, the LHCb Collaboration announced the
observation of two charmed-strange mesons $D_{s1}^*(2860)$ and
$D_{s3}^*(2860)$ with the resonance parameters
\cite{Aaij:2014xza,Aaij:2014baa}:
\begin{eqnarray*}
m_{D_{s1}^*(2860)}&=&(2859\pm12\pm6\pm23)\,\, {\mathrm{MeV}},\\
\Gamma_{D_{s1}^*(2860)}&=&(159\pm23\pm27\pm72)\,\, {\mathrm{MeV}},\\
m_{D_{s3}^*(2860)}&=&(2860.5\pm2.6\pm2.5\pm6.0)\,\,{\mathrm{MeV}},\\
\Gamma_{D_{s3}^*(2860)}&=&(53\pm7\pm4\pm6)\,\, {\mathrm{MeV}}.
\end{eqnarray*}
In addition, LHCb specified that it is the first time to identify a
spin-3 resonance $D_{s3}^*(2860)$ \cite{Aaij:2014xza,Aaij:2014baa}.
At present, the charmed-strange meson family is becoming more and
more abundant with the experimental progress.

Until now, there are good candidates of the $1S$ and $1P$ states in
the charmed-strange meson family \cite{pdg}. These newly observed
charmed-strange mesons provide a good platform to study the
properties of the higher radial and orbital excitations of the
charmed-strange meson. For example, in Ref. \cite{Sun:2009tg} Sun
and Liu suggested that $D_{sJ}(3040)$ can be a good candidate of the
$2P$ state in the charmed-strange meson family, which is the radial
excitation of $D_{s1}(2460)$. The recently reported $D_{s1}^*(2860)$
and $D_{s3}^*(2860)$ states stimulated extensive discussions of
whether they can be categorized into the $1D$ charmed-strange mesons
\cite{Song:2014mha,Wang:2014jua,Godfrey:2014fga}. In Ref.
\cite{Song:2014mha}, the two-body strong decays of $D_{s1}^*(2860)$
and $D_{s3}^*(2860)$ as the $1^3D_1$ and $1^3D_3$ states in
charmed-strange meson family were studied by the quark pair creation
model, which shows that $D_{s1}^*(2860)$ and $D_{s3}^*(2860)$ are the
$1^3D_1$ and $1^3D_3$ states, respectively. Later, Wang studied
$D_{s1}^*(2860)$ and $D_{s3}^*(2860)$ using the effective Lagrangian
approach \cite{Wang:2014jua}. Recently, Godfrey and Moats
\cite{Godfrey:2014fga} indicated that $D_{s1}^*(2860)$ and
$D_{s3}^*(2860)$ are the $1^3D_1$ and $1^3D_3$ charmed-strange mesons,
respectively. Thus, the results in Refs.
\cite{Wang:2014jua,Godfrey:2014fga} supports the assignment of
$D_{s1}^*(2860)$ and $D_{s3}^*(2860)$ proposed in Ref.
\cite{Song:2014mha}.

In this paper we shall use the method of QCD sum rule to study the
$D$-wave heavy meson doublets $(1^-,2^-)$ and $(2^-,3^-)$ containing
one heavy anti-quark and one strange
quark~\cite{Shifman:1978bx,Reinders:1984sr}. We shall work in the
framework of the heavy quark effective theory
(HQET)~\cite{Grinstein:1990mj,Eichten:1989zv,Falk:1990yz}, which has
been successful to study heavy hadrons containing a single heavy
quark. The mass of the ground state heavy mesons was studied in
Refs.~\cite{Bagan:1991sg,Neubert:1991sp,Neubert:1993mb,Broadhurst:1991fc,Ball:1993xv,Huang:1994zj}.
The masses of the lowest excited nonstrange heavy meson doublets
$(0^+,1^+)$ and $(1^+,2^+)$ were studied in
Refs.~\cite{Dai:1993kt,Dai:1996qx,Colangelo:1998ga}. The masses of
the lowest excited $\bar c s$ heavy mesons in the $(0^+,1^+)$ and
$(1^+,2^+)$ doublets were studied in Ref.~\cite{Dai:2003yg}. There
were also some early studies using the method of QCD sum rules but
in full QCD~\cite{Colangelo:1991ug,Colangelo:1992kc}. In this paper
we shall follow the procedures used in
Refs.~\cite{Dai:1993kt,Dai:1996yw,Dai:1996qx,Dai:2003yg}, and study
the $D$-wave $\bar c s$ heavy meson in the $(1^-,2^-)$ and
$(2^-,3^-)$ doublets. We shall also follow
Refs.~\cite{Dai:1993kt,Dai:1996yw,Dai:1996qx,Dai:2003yg} and
consider the ${\mathcal O}(1/m_Q)$ corrections, where $m_Q$ is the
heavy quark mass.

This paper is organized as follows. In Sec.~\ref{sec:leading}, we
introduce the interpolating currents for the $D$-wave $\bar c s$
heavy meson doublets $(1^-,2^-)$ and $(2^-,3^-)$, and use them to
perform QCD sum rule analyses at the leading order. Then in
Sec.~\ref{sec:nexttoleading} we calculate the ${\mathcal O}(1/m_Q)$
corrections. The results are summarized and discussed in
Sec.~\ref{sec:summary}.

\section{The Sum Rules at the Leading Order (in the $m_Q \rightarrow \infty$ limit)}
\label{sec:leading}

The interpolating currents for the heavy mesons with arbitrary spin
and parity have been studied and given in
Refs.~\cite{Dai:1993kt,Dai:1996yw,Dai:1996qx}. Here we briefly discuss
how we obtain the interpolating currents coupling to the $D$-wave $\bar c s$ heavy meson doublets $(1^-,2^-)$
and $(2^-,3^-)$. We use $J_{j,P,j_l}$ to denote these fields where the first two subscripts denote the spin and parity of the heavy mesons, and
the last subscript denotes the angular momentum of the light
components. We also use the following notations: $\gamma_t^\mu = \gamma^\mu - v\!\!\!\slash v^\mu$, $D^\mu_t =
D^\mu - (D \cdot v) v^\mu$, with $D^\mu = \partial^\mu - i g A^\mu$ is the gauge-covariant derivative, $h_v$ is the heavy quark field in
HQET, $v$ is the velocity of the heavy quark, and
$g_t^{\alpha_1\alpha_2}=g^{\alpha_1\alpha_2} - v^{\alpha_1}
v^{\alpha_2}$ is the transverse metric tensor.

Based on the pseudoscalar current $\bar h_v \gamma_5 q$ of $J^P=0^-$ and the vector current $\bar h_v \gamma_\mu q$ of $J^P=1^-$, we can construct the $D$-wave interpolating currents, by adding two extra derivatives. Assuming these two derivatives are both acting on the light quark, it can have either $j_l^{P_l} = 3/2^+$:
\begin{eqnarray}
\mathcal{D}_t^{\alpha} \mathcal{D}_t^{\beta} \times \gamma_\beta \gamma_5 q \, ,
\end{eqnarray}
or $j_l^{P_l} = 5/2^+$:
\begin{eqnarray}
\mathcal{D}_t^{\alpha_1} \mathcal{D}_t^{\alpha_2} \times q \, .
\end{eqnarray}
Use the light quark of $j_l^{P_l} = 3/2^+$, we can construct the interpolating currents coupling to the $D$-wave
$(1^-,2^-)$ spin doublet. One of them has the pseudoscalar structure, while the other has the vector structure:
\begin{eqnarray}
J^{\dag \alpha}_{x,-,3/2} &=& \bar h_v \gamma_5 \times \mathcal{D}_t^{\alpha} \mathcal{D}_t^{\beta} \times \gamma_\beta \gamma_5 q \, ,
\\ J^{\dag \alpha_1 \alpha_2}_{y,-,3/2} &=& \bar h_v \gamma_t^{\alpha_2} \times \mathcal{D}_t^{\alpha_1} \mathcal{D}_t^{\beta} \times \gamma_\beta \gamma_5 q \, .
\end{eqnarray}
However, they are not pure $1^-$ or $2^-$ (Particularly, $J^{\dag \alpha_1 \alpha_2}_{y,-,3/2}$ contains both $1^-$ and $2^-$ components).
Therefore, we need to do an extra process to project out their $1^-$ and $2^-$ components:
\begin{eqnarray}
J^{\dag \alpha}_{1,-,3/2} &=& \sqrt{\frac{3}{4}} \bar h_v (-i) \big
( \mathcal{D}_t^{\alpha} - \frac{1}{3} \gamma_t^{\alpha} \slashed
D_t \big ) \slashed D_t q \, , \label{eq:current1}
\\ J^{\dag \alpha_1 \alpha_2}_{2,-,3/2} &=& \sqrt{\frac{1}{2}} \bar h_v \gamma^5 \frac{(-i)^2}{2} \big ( \gamma_t^{\alpha_1} \mathcal{D}_t^{\alpha_2}\slashed D_t + \gamma_t^{\alpha_2} \mathcal{D}_t^{\alpha_1}\slashed D_t \nonumber\\&&- \frac{2}{3}g_t^{\alpha_1 \alpha_2} \mathcal{D}_t \cdot \mathcal{D}_t \big ) q \, ,
\label{eq:current2}
\end{eqnarray}
where we have modified their expressions to be consistent with Refs.~\cite{Dai:1993kt,Dai:1996yw,Dai:1996qx}.
Similarly, we can construct the following interpolating currents to study the $D$-wave
$(2^-,3^-)$ spin doublet:
\begin{eqnarray}
J^{\dag \alpha_1 \alpha_2}_{2,-,5/2} &=& \sqrt{\frac{5}{6}} \bar h_v
\gamma^5 \frac{(-i)^2}{2} \big (\mathcal{D}_t^{\alpha_2}
\mathcal{D}_t^{\alpha_1} +
\mathcal{D}_t^{\alpha_1}\mathcal{D}_t^{\alpha_2} - \frac{2}{5}
\mathcal{D}_t^{\alpha_2}\gamma_t^{\alpha_1}\slashed D_t
\nonumber\\&&- \frac{2}{5}
\mathcal{D}_t^{\alpha_1}\gamma_t^{\alpha_2}\slashed
D_t-\frac{2}{5}g_t^{\alpha_1 \alpha_2} \mathcal{D}_t \cdot
\mathcal{D}_t \big ) q \, , \label{eq:current3}
\\ J^{\dag \alpha_1 \alpha_2 \alpha_3}_{3,-,\frac{5}{2}} &=&\sqrt{\frac{1}{2}} \bar h_v \mathcal{S}_1 [\gamma_t^{\alpha_1} (-i)^2 \mathcal{D}_t^{\alpha_2} \mathcal{D}_t^{\alpha_3}] q \, ,
\label{eq:current4}
\end{eqnarray}
where $\mathcal{S}_1$
denotes symmetrization and subtracting the trace terms in the sets
$(\alpha_1 \alpha_2 \alpha_3)$.

Based on the scalar current $\bar h_v q$ of $J^P=0^+$ and the axial-vector current $\bar h_v \gamma_\mu \gamma_5 q$ of $J^P=1^+$, we can also construct the $D$-wave interpolating currents, by adding only one derivative. For example, the light quark containing one derivative can have $j_l^{P_l} = 3/2^-$:
\begin{eqnarray}
\mathcal{D}_t^{\alpha} \times q \, .
\end{eqnarray}
Then it can be used to construct the interpolating currents coupling to the $D$-wave $(1^-,2^-)$ spin doublet:
\begin{eqnarray}
J^{\prime \dag \alpha}_{x,-,3/2} &=& \bar h_v \times \mathcal{D}_t^{\alpha} \times q \, ,
\\ J^{\prime \dag \alpha_1 \alpha_2}_{y,-,3/2} &=& \bar h_v \gamma_t^{\alpha_2} \gamma_5 \times \mathcal{D}_t^{\alpha_1} \times q \, .
\end{eqnarray}
These two currents can be further manipulated to obtain the pure $1^-$ and $2^-$ currents.
\begin{eqnarray}
J^{\prime \dag \alpha}_{1,-,3/2} &=& \sqrt{\frac{3}{4}} \bar h_v (-i) \big
( \mathcal{D}_t^{\alpha} - \frac{1}{3} \gamma_t^{\alpha} \slashed
D_t \big ) q \, ,
\\ J^{\prime \dag \alpha_1 \alpha_2}_{2,-,3/2} &=& \sqrt{\frac{1}{2}} \bar h_v \gamma^5 \frac{(-i)^2}{2} \big ( \gamma_t^{\alpha_1} \mathcal{D}_t^{\alpha_2} + \gamma_t^{\alpha_2} \mathcal{D}_t^{\alpha_1} \nonumber\\&&- \frac{2}{3}g_t^{\alpha_1 \alpha_2} \slashed D_t \big ) q \, .
\end{eqnarray}
However, compared with these currents containing one derivative, the currents containing
two derivatives seem to describe the internal structure of the $D$-wave heavy mesons in a more appropriate way, so
we shall use Eqs.~(\ref{eq:current1}), (\ref{eq:current2}), (\ref{eq:current3}) and (\ref{eq:current4}) to perform QCD sum rule analyses.
Moreover, the two currents $J_{1,-,3/2}$ and $J_{2,-,3/2}$ give
identical sum rules at the leading order, and the sum rules at the
$O(1/m_Q)$ order can be obtained using either of
them~\cite{Dai:1993kt,Dai:1996yw,Dai:1996qx,Dai:2003yg}
(ideally the results should be identical, while actually they have
small differences but negligible).
Accordingly, we only need to use one of them to perform QCD sum rule
analyses. So do $J_{2,-,5/2}$ and $J_{3,-,5/2}$. In the following discussions we shall
use $J_{1,-,3/2}$ and $J_{3,-,5/2}$, because they couple to the newly
observed $D_{s1}^*(2860)$ and $D_{s3}^*(2860)$ states by LHCb~\cite{Aaij:2014xza,Aaij:2014baa}.

In the $m_Q \rightarrow \infty$ limit we can assume
$|j,P,j_l\rangle$ to be the heavy meson state with the quantum
numbers $j, P$ and $j_l$, and the relation between this state and
the relevant interpolating field is
\begin{eqnarray}
\langle 0| J^{\alpha_1\cdots\alpha_j}_{j,P,j_l}
|j^\prime,P^\prime,j_l^\prime \rangle = f_{P,j_l} \delta_{j
j^\prime} \delta_{P P^\prime} \delta_{j_l j_l^\prime}
\eta^{\alpha_1\cdots\alpha_j}_t \, ,
\end{eqnarray}
where $f_{P, j_l}$ is the decay constant. It has the same value for
the two states in the same doublet in the $m_Q \rightarrow \infty$
limit. $\eta^{\alpha_1\cdots\alpha_j}_t$ is the transverse,
symmetric, and traceless polarization tensor. In this paper we need
to use $\eta^{\alpha}_t$ and $\eta^{\alpha_1\alpha_2\alpha_3}_t$ which have the following property
at the leading order
\begin{eqnarray}
\eta^{\alpha}_t \eta^{*\beta}_t &=& \tilde g^{\alpha \beta}_t \, ,
\\ \eta^{\alpha_1\alpha_2\alpha_3}_t \eta^{*\beta_1\beta_2\beta_3}_t &=& \mathcal{S}_2
[\tilde g^{\alpha_1 \beta_1}_t \tilde g^{\alpha_2 \beta_2}_t \tilde g^{\alpha_3 \beta_3}_t] \, ,
\end{eqnarray}
where $\tilde g^{\mu \nu} = g^{\mu \nu} - q^\mu q^\nu / m^2$ and $\mathcal{S}_2$ denotes symmetrization and subtracting the trace
terms in the sets $(\alpha_1 \alpha_2 \alpha_3)$ and $(\beta_1 \beta_2
\beta_3)$.

Using the two interpolating currents $J_{1,-,3/2}$ and
$J_{3,-,5/2}$, we can construct the two-point correlation function
\begin{eqnarray}
\Pi^{\alpha_1\cdots\alpha_j,\beta_1\cdots\beta_j}_{j,P,j_l} (\omega)
&=& i \int d^4 x e^{i k x} \langle 0 |
T[J^{\alpha_1\cdots\alpha_j}_{j,P,j_l}(x)
J^{\dagger\beta_1\cdots\beta_j}_{j,P,j_l}(0)] | 0 \rangle\nonumber
\\  &=& (-1)^j \mathcal{S}_3 [ g_t^{\alpha_1 \beta_1} \cdots g_t^{\alpha_j \beta_j} ] \Pi_{j,P,j_l} (\omega) \, ,\label{eq:pi}
\end{eqnarray}
where $\omega = 2 v \cdot k$ is twice the external off-shell energy,
and $\mathcal{S}_3$ denotes symmetrization and subtracting the trace
terms in the sets $(\alpha_1 \cdots \alpha_j)$ and $(\beta_1 \cdots
\beta_j)$. At the hadron level, it can be written as
\begin{eqnarray}
\Pi_{j,P,j_l}(\omega) = {f_{P,j_l}^2 \over 2 \bar \Lambda_{j,P,j_l}
- \omega} + \mbox{higher states} \, , \label{eq:pole}
\end{eqnarray}
where $\bar \Lambda_{j,P,j_l} = \lim_{m_Q \rightarrow \infty}
(m_{j,P,j_l} - m_Q)$, and $m_{j,P,j_l}$ is the mass of the
lowest-lying heavy meson state which
$J^{\alpha_1\cdots\alpha_j}_{j,P,j_l}(x)$ couples to. At the quark
and gluon level, we can calculate the two-point correlation function
(\ref{eq:pi}) using the method of QCD sum rule. To do this we follow
the approaches used in
Refs.~\cite{Dai:1993kt,Dai:1996yw,Dai:1996qx,Dai:2003yg}. After
inserting Eq.~(\ref{eq:current1}) and (\ref{eq:current4}) into
Eq.~(\ref{eq:pi}), and performing the Borel transformation, we
obtain
\begin{eqnarray}
&&\Pi_{1,-,3/2}(\omega_c, T) = f_{-,3/2}^2 e^{-2 \bar
\Lambda_{1,-,3/2} / T} \nonumber\\  \quad&&= {7 \over 2560 \pi^2}
\int_{2 m_s}^{\omega_c} [ \omega^6 + 2 m_s \omega^5 - 10 m_s^2
\omega^4 ] e^{-\omega/T} d\omega\nonumber\\&& \quad - {1 \over 8
\pi} {\langle \alpha_s GG \rangle} T^3 \, ,\label{eq:ope1}
\\ &&\Pi_{3,-,5/2}(\omega_c, T) = f_{-,5/2}^2 e^{-2 \bar \Lambda_{3,-,5/2} / T}
\nonumber \\ \quad&&={1 \over 640 \pi^2} \int_{2 m_s}^{\omega_c} [
\omega^6 + 2 m_s \omega^5 - 10 m_s^2 \omega^4 ] e^{-\omega/T}
d\omega\nonumber\\&&\quad - {3 \over 32 \pi} {\langle \alpha_s GG
\rangle} T^3 \, .\label{eq:ope2}
\end{eqnarray}
We note that there are $2 \times 2 = 4$ derivatives, and so the
calculations are not easy. To deal with them, we have used a
software called {\it Mathematica} with a package called
$FeynCalc$~\cite{feyncalc}. Moreover, we do not consider
the radiative corrections in order to simply our calculations.

\begin{figure}[hbt]
\begin{center}
\scalebox{0.6}{\includegraphics{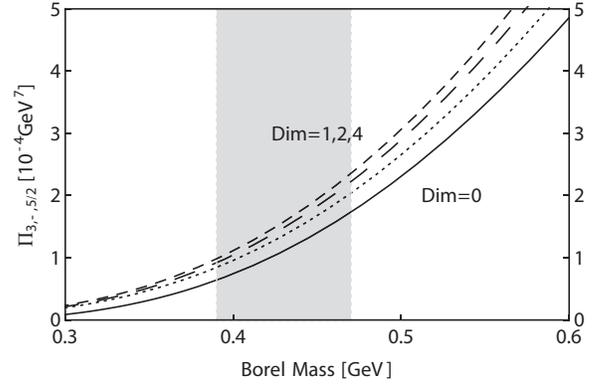}} \caption{Various
contributions to Eq.~(\ref{eq:ope2}), as functions of the Borel mass
$M_B$ in units of GeV$^{7}$ at $\omega_c$ = 2.7 GeV. The labels
indicate the dimension up to which the OPE terms are included.}
\label{fig:pi}
\end{center}
\end{figure}

Particularly, the quark condensate $\langle \bar q q \rangle$ and
the mixed condensate $\langle g_s \bar q \sigma G q \rangle$ both
vanish in this case. This is much different from those sum rules for
$(0^+, 1^+)$ and $(1^+, 2^+)$
doublets~\cite{Dai:1993kt,Dai:1996yw,Dai:1996qx,Dai:2003yg}, and it
makes the convergence of Eq.~(\ref{eq:ope1}) and (\ref{eq:ope2})
very good. To clearly see this, we show the convergence of
Eq.~(\ref{eq:ope2}) in Fig.~\ref{fig:pi}, where $\omega_c$ is taken
to be $2.7$ GeV, and the following values for the gluon condensate and
the strange quark mass are
used~\cite{Dai:1993kt,Dai:1996yw,Dai:1996qx,Dai:2003yg,Ioffe:2005ym}:
%
\begin{eqnarray}
&&\langle {\alpha_s\over\pi} GG\rangle = 0.005 \pm 0.004 \mbox{ GeV}^4\, ,
\\ \label{condensates} && m_s = 0.15 \mbox{ GeV} \, .
\end{eqnarray}

Finally, we differentiate Eq.~(\ref{eq:ope1}) and (\ref{eq:ope2})
with respect to $-2 / T$, divide the results by themselves, and
obtain
\begin{equation}
\Lambda_{j,P,j_l}(\omega_c, T) =
\frac{\frac{\partial}{\partial(-2/T)}\Pi_{j,P,j_l}(\omega_c,
T)}{\Pi_{j,P,j_l}(\omega_c, T)} \, . \label{eq:mass}
\end{equation}
The results can be furtherly used to evaluate $f_{P,j_l}$:
\begin{equation}
f_{P,j_l}(\omega_c, T) = \sqrt{\Pi_{j,P,j_l}(\omega_c, T) \times
e^{2 \bar \Lambda_{j,P,j_l}(\omega_c, T) / T}} \, .
\label{eq:coupling}
\end{equation}
Here we note again that the sum rule obtained by using the other
current $J_{2,-,3/2}$ is very similar to Eq.~(\ref{eq:ope1}). Hence, we have $\Lambda_{1,-,3/2}
= \Lambda_{2,-,3/2}$, and we shall use another symbol
$\Lambda_{-,3/2}$ to denote them. Similarly we shall use the symbol
$\Lambda_{-,5/2}$ to denote $\Lambda_{2,-,5/2}$ and
$\Lambda_{3,-,5/2}$. To differentiate the masses within the same
doublet, we need to work at the $O(1/m_Q)$ order, which will be done
in the next section.

\begin{figure}[hbt]
\begin{center}
\scalebox{0.6}{\includegraphics{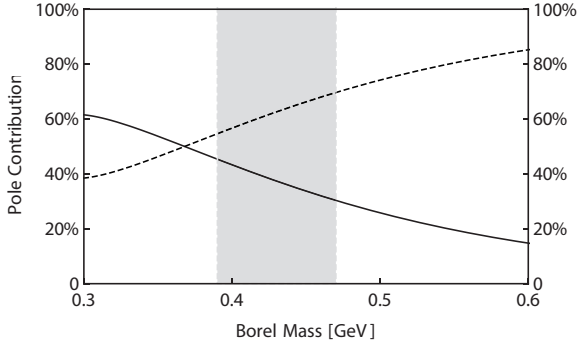}} \caption{The solid curve
shows the pole contribution and the dashed curve shows the continuum
contribution ($= 1 - $pole contribution), when $J_{3,-,5/2}$ is used
and $\omega_c$ is fixed to be 2.7 GeV. } \label{fig:pole}
\end{center}
\end{figure}

To perform the numerical analysis, firstly we require that the
high-order power corrections be less than 30\% of the perturbation
term, and obtain the minimum value $T_{min}$ of the allowed Borel
parameter; secondly we require that the pole contribution
%
\begin{equation}
\label{eq_pole} \mbox{Pole contribution} \equiv \frac{
\Pi_{j,P,j_l}(\omega_c, T) }{ \Pi_{j,P,j_l}( \infty , T) } \, ,
\end{equation}
%
is larger than 30\%, and obtain the maximum value $T_{max}$ of the
allowed Borel parameter. Altogether we have the working interval
$T_{min}<T<T_{max}$ for a fixed $\omega_c$. In the sum rules
(\ref{eq:ope1}) and (\ref{eq:ope2}) $\omega_c$ is free parameter,
and we choose it to be around 2.7 GeV for both $\Pi_{1,-,3/2}$ and
$\Pi_{3,-,5/2}$, because the $D_{s1}^*(2860)$ and $D_{s3}^*(2860)$
observed by LHCb have similar masses $2859$ MeV and $2860.5$
MeV~\cite{Aaij:2014xza}. Using this $\omega_c = 2.7$ GeV, we obtain
our working regions, around $0.35$ GeV $< T < 0.48$ GeV for
$J_{1,-,3/2}$, and around $0.39$ GeV $< T < 0.47$ GeV for
$J_{3,-,5/2}$. As an example, we show the comparison between the
pole and continuum contributions for $J_{3,-,5/2}$ in
Fig.~\ref{fig:pole}.

\begin{figure}[hbt]
\begin{center}
\scalebox{0.6}{\includegraphics{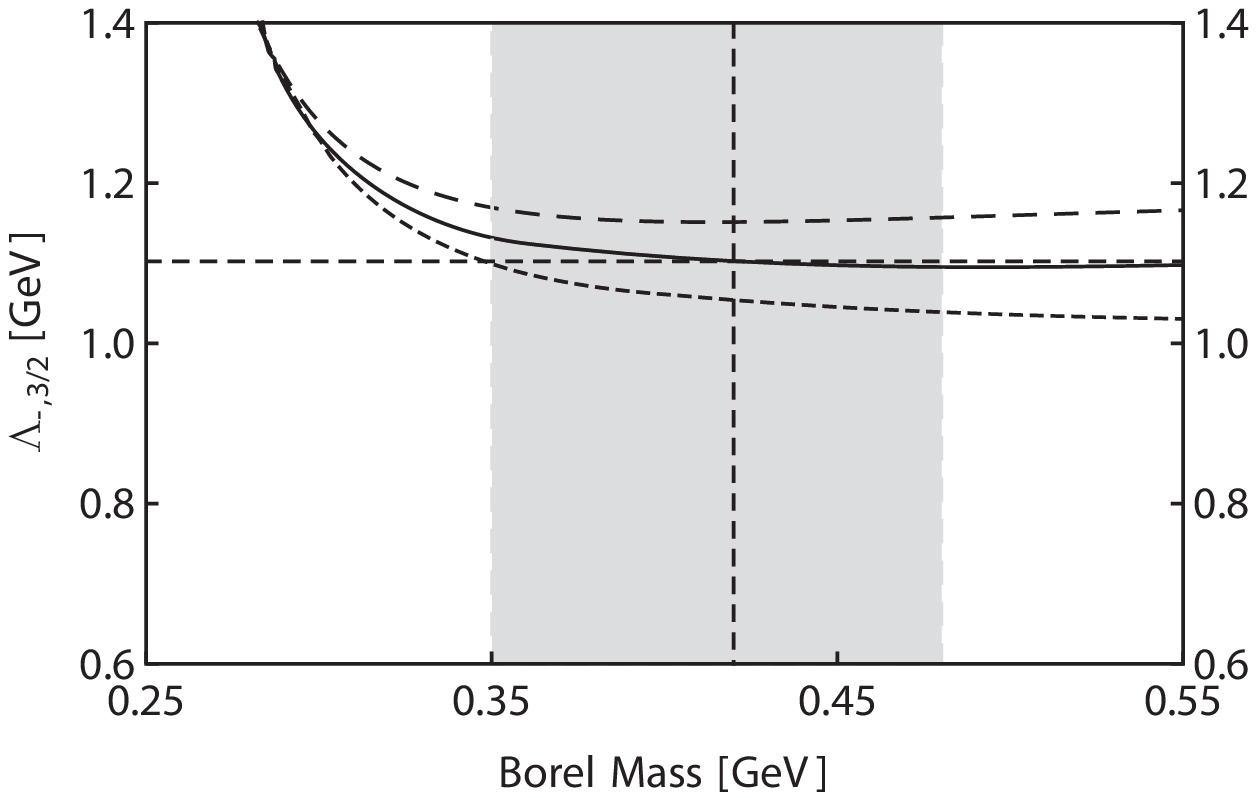}}
\scalebox{0.6}{\includegraphics{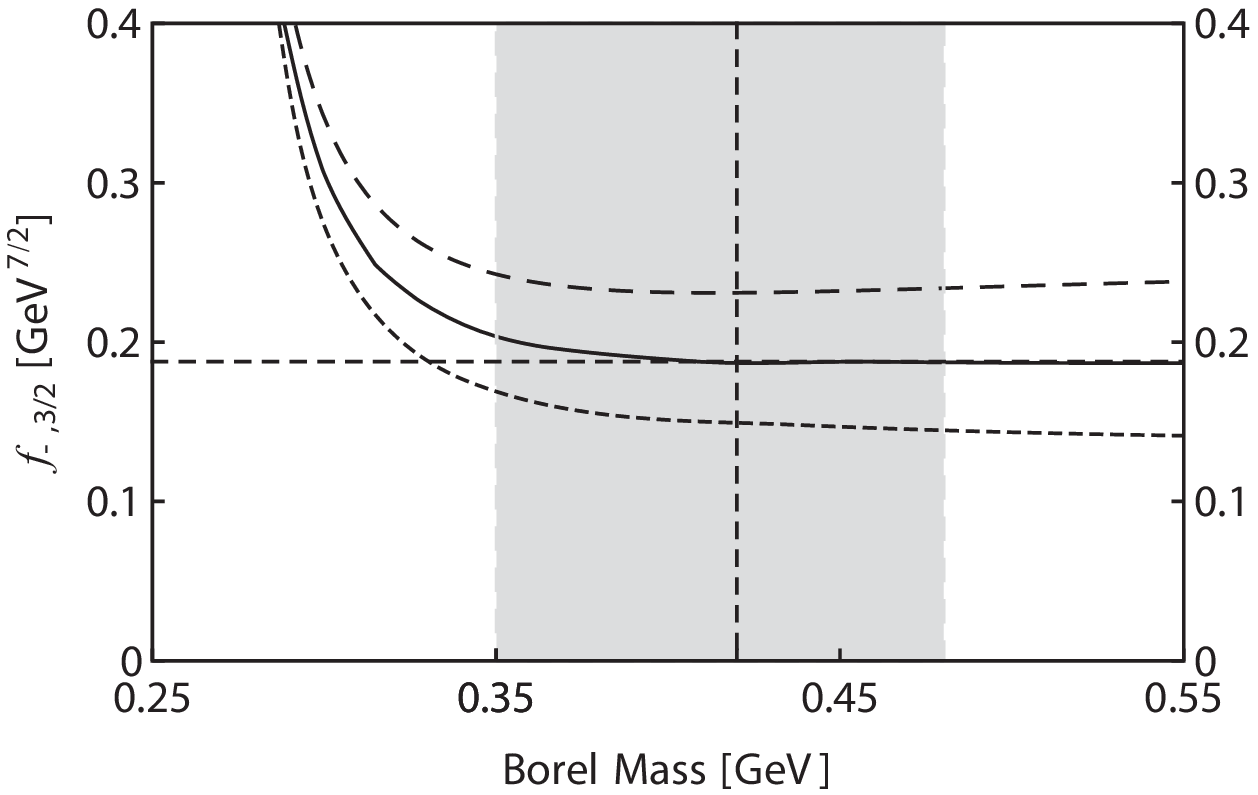}} \caption{The
variation of $\Lambda_{-,3/2}$ and $f_{-,3/2}$ with respect to the
Borel mass $T$ and the threshold value $\omega_c$. The short-dashed,
solid and long-dashed curves are obtained by fixing $\omega_c =
2.5$, 2.7 and 2.9 GeV, respectively. Our working region is $0.35$
GeV $< T < 0.48$ GeV.} \label{fig:leading1}
\end{center}
\end{figure}

Finally, we solve Eq.~(\ref{eq:mass}) and Eq.~(\ref{eq:coupling}),
and evaluate $\Lambda_{-,j_l}$ and $f_{-,j_l}$. We show the
variations of $\Lambda_{-,3/2}$ and $f_{-,3/2}$ with respect to the
Borel mass $T$ and the threshold value $\omega_c$ in
Fig.~\ref{fig:leading1}. These figures are shown in the region
$0.25$ GeV $< T < 0.55$ GeV, but we find that their
dependence on the Borel mass $T$ becomes weaker in our working
region $0.35$ GeV $< T < 0.48$ GeV. We obtain the following
numerical results:
\begin{eqnarray}
\Lambda_{-,3/2} &=& 1.10 \pm 0.06 \mbox{ GeV} \, ,
\\ f_{-,3/2} &=& 0.19 \pm 0.05 \mbox{ GeV}^{7/2} \, ,
\end{eqnarray}
where the central value corresponds to $T=0.42$ GeV and $\omega_c =
2.7$ GeV.

\begin{figure}[hbt]
\begin{center}
\scalebox{0.6}{\includegraphics{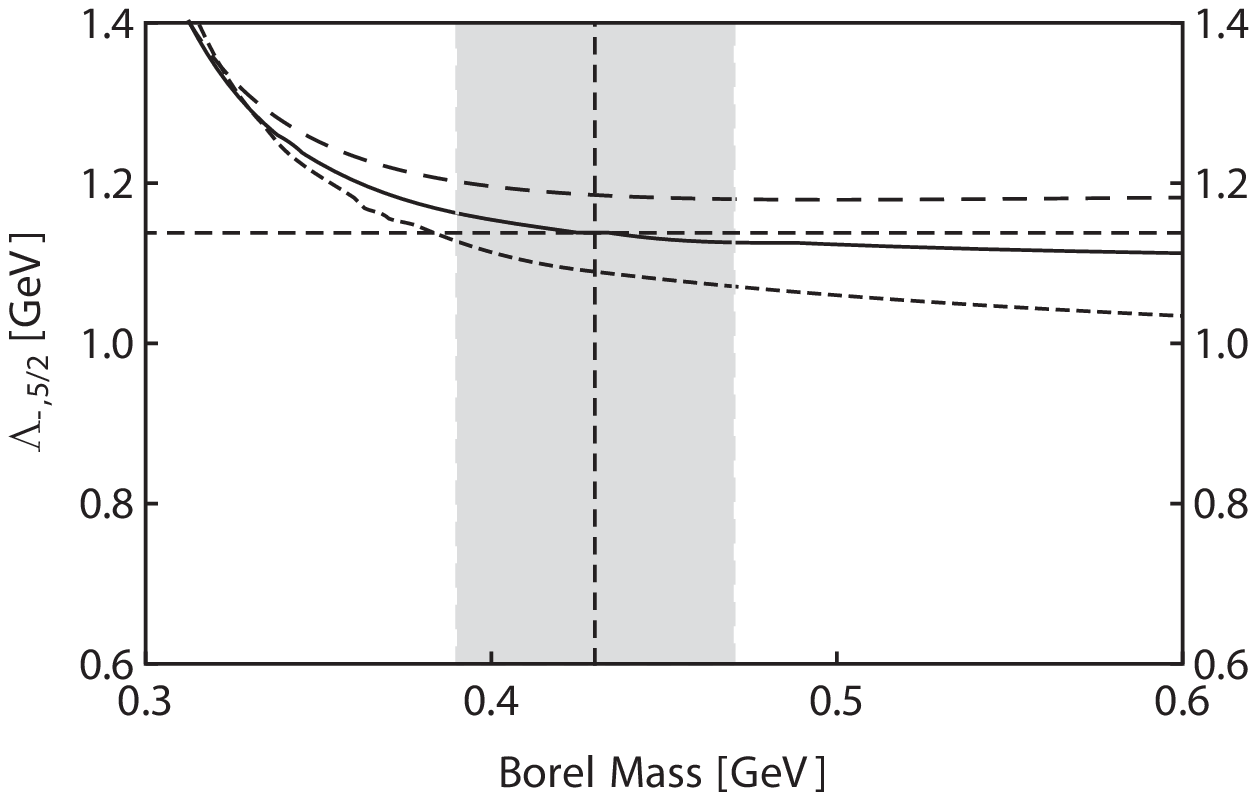}}
\scalebox{0.6}{\includegraphics{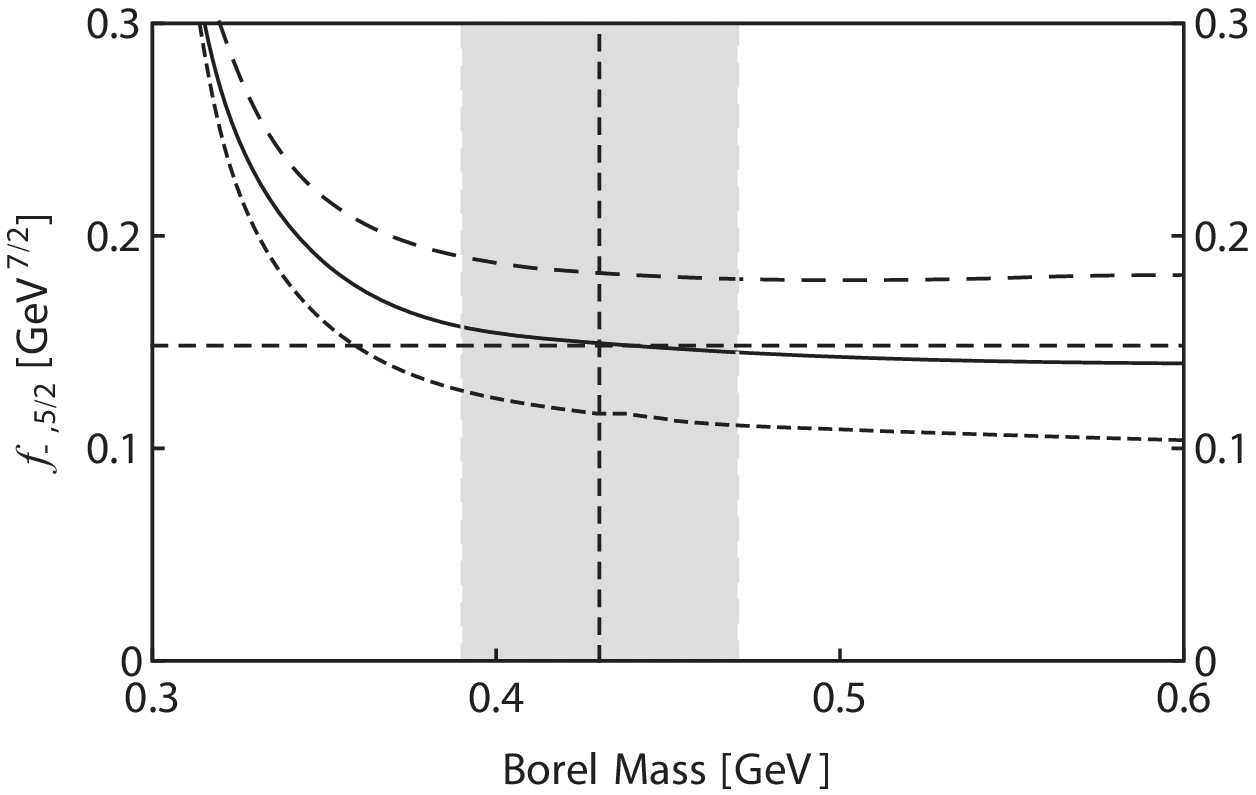}} \caption{The
variation of $\Lambda_{-,5/2}$ and $f_{-,5/2}$ with respect to the
Borel mass $T$ and the threshold value $\omega_c$. The short-dashed,
solid and long-dashed curves are obtained by fixing $\omega_c =
2.5$, 2.7 and 2.9 GeV, respectively. Our working region is $0.39$
GeV $< T < 0.47$ GeV.} \label{fig:leading2}
\end{center}
\end{figure}

Similarly, we show the variations of $\Lambda_{-,5/2}$ and
$f_{-,5/2}$ in Fig.~\ref{fig:leading2}. These figures are shown in
the region $0.30$ GeV $< T < 0.60$ GeV. Again we find that their
dependence on the Borel mass $T$ becomes weaker in our working
region $0.39$ GeV $< T < 0.47$ GeV. We obtain the following
numerical results:
\begin{eqnarray}
\Lambda_{-,5/2} &=& 1.14 \pm 0.05 \mbox{ GeV} \, ,
\\ f_{-,5/2} &=& 0.15 \pm 0.04 \mbox{ GeV}^{7/2} \, ,
\end{eqnarray}
where the central value corresponds to $T=0.43$ GeV and $\omega_c =
2.7$ GeV.

\section{The Sum Rules at the ${\mathcal O}(1/m_Q)$ Order}
\label{sec:nexttoleading}

The Lagrangian of HQET, up to the ${\mathcal O}(1/m_Q)$ order, can
be written as~\cite{Dai:1996qx,Dai:2003yg}
\begin{eqnarray}
\mathcal{L}_{\rm eff} = \overline{h}_{v}iv\cdot Dh_{v} +
\frac{1}{2m_{Q}}\mathcal{K} + \frac{1}{2m_{Q}}\mathcal{S} \, ,
\label{eq:next}
\end{eqnarray}
where $\mathcal{K}$ is the operator of the nonrelativistic kinetic
energy with a negative sign:
\begin{eqnarray}
\mathcal{K} = \overline{h}_{v}(iD_{t})^{2}h_{v} \, ,
\end{eqnarray}
and $\mathcal S$ is the Pauli term to describe the chromomagnetic
interaction:
\begin{eqnarray}
\mathcal{S}= \frac{g}{2} C_{mag} (m_{Q}/\mu) \overline{h}_{v}
\sigma_{\mu\nu} G^{\mu\nu} h_{v} \, ,
\end{eqnarray}
where $C_{mag} (m_{Q}/\mu) = [ \alpha_s(m_Q) / \alpha_s(\mu)
]^{3/\beta_0}$ and $\beta_0 = 11 - 2 n_f /3$.

We use $\delta m$ and $ \delta f$ to denote the corrections to the
mass $m_{j,P,j_l}$ and the coupling constant $f_{P,j_l}$ at the
$\mathcal{O}(1/m_Q)$ order. The pole term on the hadron side,
Eq.~(\ref{eq:pole}), can be written as:
\begin{eqnarray}
\Pi(\omega)_{pole} &=& \frac{(f+\delta
f)^{2}}{2(\overline{\Lambda}+\delta m)-\omega}
 \nonumber\\ &=& \frac{f^{2}}{2\overline{\Lambda}-\omega}-\frac{2\delta mf^{2}}{(2\overline{\Lambda}-\omega)^{2}}+\frac{2f\delta f}{2\overline{\Lambda}-\omega} \, .\label{eq:correction}
\end{eqnarray}
In this paper we shall only evaluate $\delta m$. To do this, we use
the Lagrangian (\ref{eq:next}) defined at the ${\mathcal O}(1/m_Q)$
order, and consider the following three-point correlation functions
\begin{eqnarray}
&&\delta_{O}\Pi_{j,P,i}^{\alpha_{1}\cdots\alpha_{j},\beta_{1}\cdots\beta_{j}}(\omega
, \omega ') \nonumber\\\quad&&= i^{2}\int d^{4}xd^{4}ye^{ik\cdot
x-ik'\cdot y}\,\langle0|T[J_{j,P,i}^{\alpha_{1}\cdots
\alpha_{j}}(x)O(0)J_{j,P,i}^{\dag\beta_{1}\cdots
\beta_{j}}(y)]|0\rangle \label{eq:nextpi} \nonumber\\ \quad&&=
(-1)^j \mathcal{S}_2 [ g_t^{\alpha_1 \beta_1} \cdots g_t^{\alpha_j
\beta_j} ] \delta_{O} \Pi_{j,P,j_l} (\omega) \, ,
\end{eqnarray}
where $O = \mathcal{K}$ or $\mathcal{S}$. At the hadron level, we
can pick their pole parts
\begin{eqnarray}
\delta_{\mathcal{K}}\Pi(\omega,\omega')_{j,P,j_l} &=&
\frac{f^{2}K_{P,j_{l}}}{(2\overline{\Lambda}-\omega)(2\overline{\Lambda}-\omega')}
+\frac{f^{2}G_{\mathcal{K}}(\omega')}{2\overline{\Lambda}-\omega}
\nonumber\\&&+\frac{f^{2}G_{\mathcal{K}}(\omega)}{2\overline{\Lambda}-\omega'}
\, , \label{eq:K}
\\ \delta_{\mathcal{S}}\Pi(\omega,\omega')_{j,P,j_l} &=& \frac{d_{M}f^{2}\Sigma_{P,j_{l}}}{(2\overline{\Lambda}-\omega)(2\overline{\Lambda}-\omega')}
+\frac{d_{M}f^{2}G_{\mathcal{S}}(\omega')}{2\overline{\Lambda}-\omega}
\,
\nonumber\\&&+\frac{d_{M}f^{2}G_{\mathcal{S}}(\omega)}{2\overline{\Lambda}-\omega'}
\, , \label{eq:S}
\end{eqnarray}
where
\begin{eqnarray}
\nonumber K_{P,j_{l}} &=& \langle
j,P,j_{l}|\overline{h}_{v}(iD_{\bot})^{2}h_{v}|j,P,j_{l}\rangle \, ,
\\ \nonumber 2d_{M}\Sigma_{P,j_{l}} &=& \langle j,P,j_{l}| g \overline{h}_{v}\sigma_{\mu\nu}G^{\mu\nu}h_{v}|j,P,j_{l}\rangle \, ,
\\ d_{M} &=& d_{j,j_{l}} \, ,
\\ \nonumber d_{j_{l}-1/2,j_{l}} &=& 2j_{l}+2
\\ \nonumber d_{j_{l}+1/2,j_{l}} &=& -2j_{l} \, .
\end{eqnarray}
From these equations we know that the term $\mathcal S$ causes a
mass splitting within the same doublet, while the term $\mathcal K$
does not. Moreover, the term $\mathcal S$ can also cause a mixing of
states with the same $j,P$ but different $j_l$, such as a mass
splitting between $|2,-,3/2\rangle$ and $|2,-,5/2\rangle$. This
effect has been studied in Ref.~\cite{Dai:1998ve}, where its
corrections are found to be negligible. Hence, we do not consider
this effect in this paper.

Fixing $\omega = \omega^\prime$ and comparing
Eq.~(\ref{eq:correction}), Eq.~(\ref{eq:K}) and Eq.~(\ref{eq:S}), we
obtain
\begin{eqnarray}
\delta m_{P,j_{l}} = -\frac{1}{4m_{Q}}(K_{P,j_{l}} +
d_{M}C_{mag}\Sigma_{P,j_{l}} ) \, .
\end{eqnarray}

At the quark and gluon level, we can calculate
Eqs.~(\ref{eq:nextpi}) using the method of QCD sum rule, and
evaluate $K_{P,j_{l}}$ and $\Sigma_{P,j_{l}}$. To do this, again we
follow the approaches used in Refs.~\cite{Dai:1996qx,Dai:2003yg}:
after inserting Eq.~(\ref{eq:current1}) and (\ref{eq:current4}) into
Eq.~(\ref{eq:nextpi}), we make a double Borel transformation for
both $\omega$ and $\omega^\prime$, and obtain two Borel parameters
$T_1$ and $T_2$. Then we take these two Borel parameters to be
equal, and obtain the following two sum rules for $K_{-,3/2}$ and
$\Sigma_{-,3/2}$:
\begin{eqnarray}
&&f_{-,3/2}^2 K_{-,3/2} e^{-2 \bar \Lambda_{-,3/2} / T}
\label{eq:Kc1} \nonumber\\&&= -{11 \over 7168 \pi^2} \int_{2
m_s}^{\omega_c} \omega^8 e^{-\omega/T} d\omega + {91 \over 64 \pi}
{\langle \alpha_s GG \rangle} T^5 \, ,
\\&& f_{-,3/2}^2 \Sigma_{-,3/2} e^{-2 \bar \Lambda_{-,3/2} / T}
\label{eq:Sc1} = {7 \over 240 \pi} {\langle \alpha_s GG \rangle} T^5
\, ,
\end{eqnarray}
and the following two sum rules for $K_{-,5/2}$ and
$\Sigma_{-,5/2}$:
\begin{eqnarray}
&&f_{-,5/2}^2 K_{-,5/2} e^{-2 \bar \Lambda_{-,5/2} / T}
\label{eq:Kc2} \nonumber\\&&= -{1 \over 1280 \pi^2} \int_{2
m_s}^{\omega_c} \omega^8 e^{-\omega/T} d\omega + {71 \over 96
\pi} {\langle \alpha_s GG \rangle} T^5 \, ,
\\ &&f_{-,5/2}^2 \Sigma_{-,5/2} e^{-2 \bar \Lambda_{-,5/2} / T}
\label{eq:Sc2} = {1 \over 40 \pi} {\langle \alpha_s GG \rangle}
T^5 \, .
\end{eqnarray}
We note that in these sum rules the $m_s$ corrections are neglected.

\begin{figure}[hbt]
\begin{center}
\scalebox{0.6}{\includegraphics{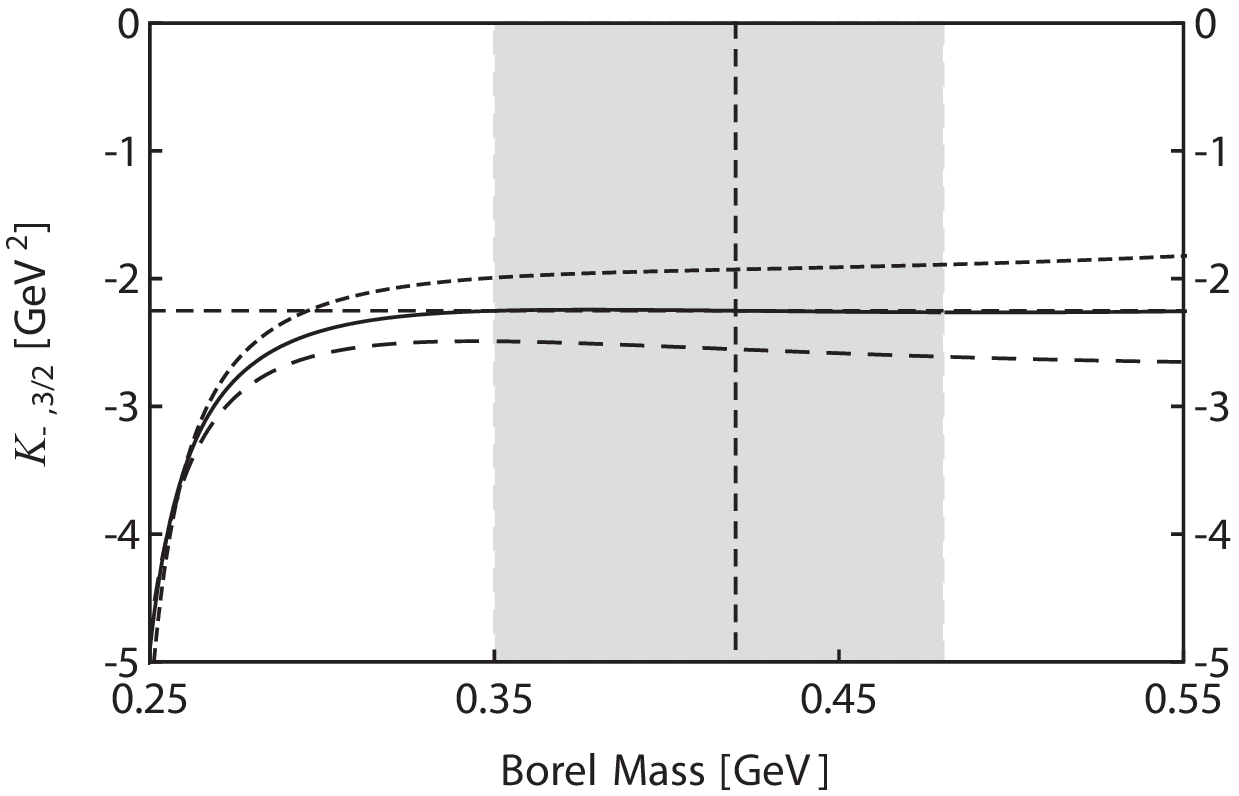}}
\scalebox{0.6}{\includegraphics{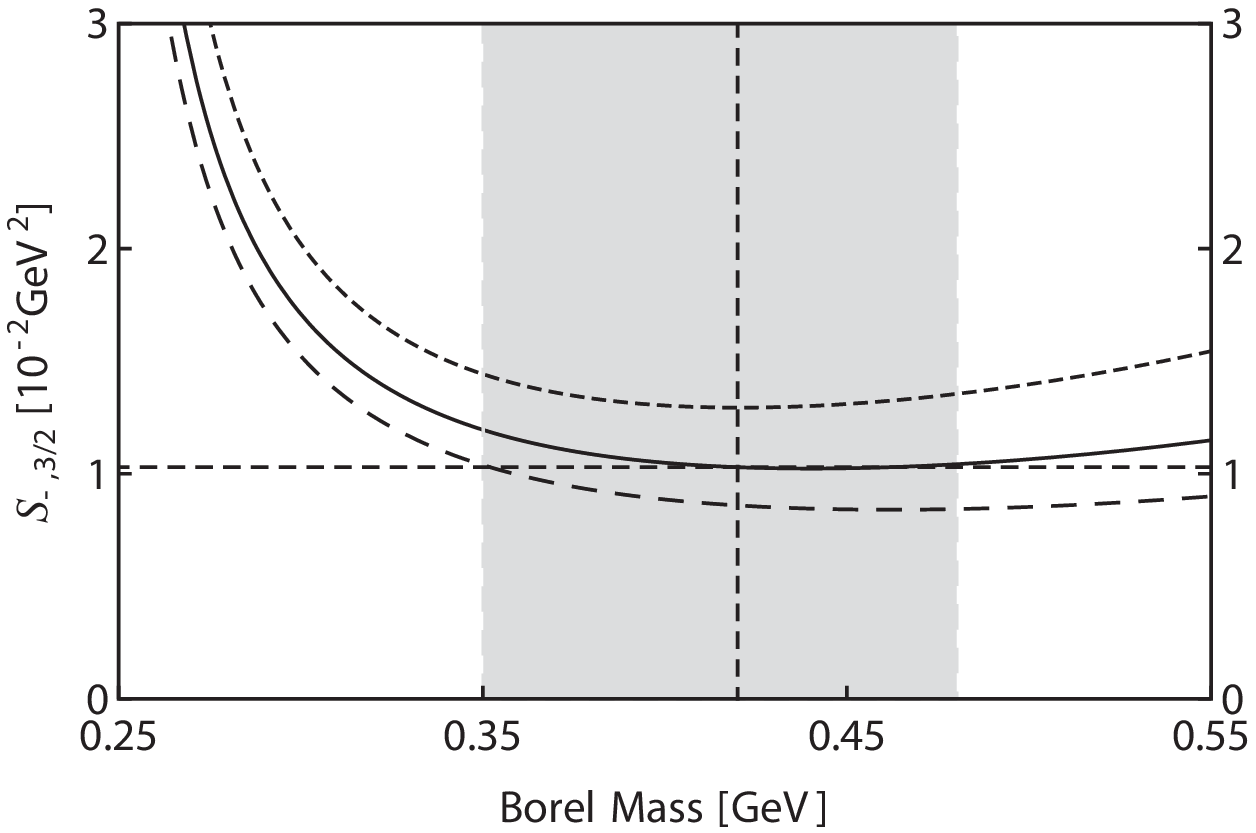}} \caption{The variation
of $K_{-,3/2}$ and $\Sigma_{-,3/2}$ with respect to the Borel mass
$T$ and the threshold value $\omega_c$. The short-dashed, solid and
long-dashed curves are obtained by fixing $\omega_c = 2.5$, 2.7 and
2.9 GeV, respectively. Our working region is $0.35$ GeV $< T < 0.48$
GeV.} \label{fig:next1}
\end{center}
\end{figure}

Finally, we obtain $K_{-,3/2}$ and $\Sigma_{-,3/2}$ by simply
dividing Eq.~(\ref{eq:Kc1}) and (\ref{eq:Sc1}) by the sum rule
(\ref{eq:ope1}), and $K_{-,5/2}$ and $\Sigma_{-,5/2}$ by simply
dividing Eq.~(\ref{eq:Kc2}) and (\ref{eq:Sc2}) by the sum rule
(\ref{eq:ope2}). We show the variations of $K_{-,3/2}$ and
$\Sigma_{-,3/2}$ with respect to the Borel mass $T$ and the
threshold value $\omega_c$ in Fig.~\ref{fig:next1} in the region
$0.25$ GeV $< T < 0.55$ GeV, and their
dependence on the Borel mass $T$ becomes weaker in our working
region $0.35$ GeV $< T < 0.48$ GeV. We obtain the following numerical
results:
\begin{eqnarray}
K_{-,3/2} &=& -2.25 \pm 0.36 \mbox{ GeV}^2 \, ,
\\ \Sigma_{-,3/2} &=& 0.010 \pm 0.004 \mbox{ GeV}^{2} \, ,
\end{eqnarray}
where the central value corresponds to $T=0.42$ GeV and $\omega_c =
2.7$ GeV.

\begin{figure}[hbt]
\begin{center}
\scalebox{0.6}{\includegraphics{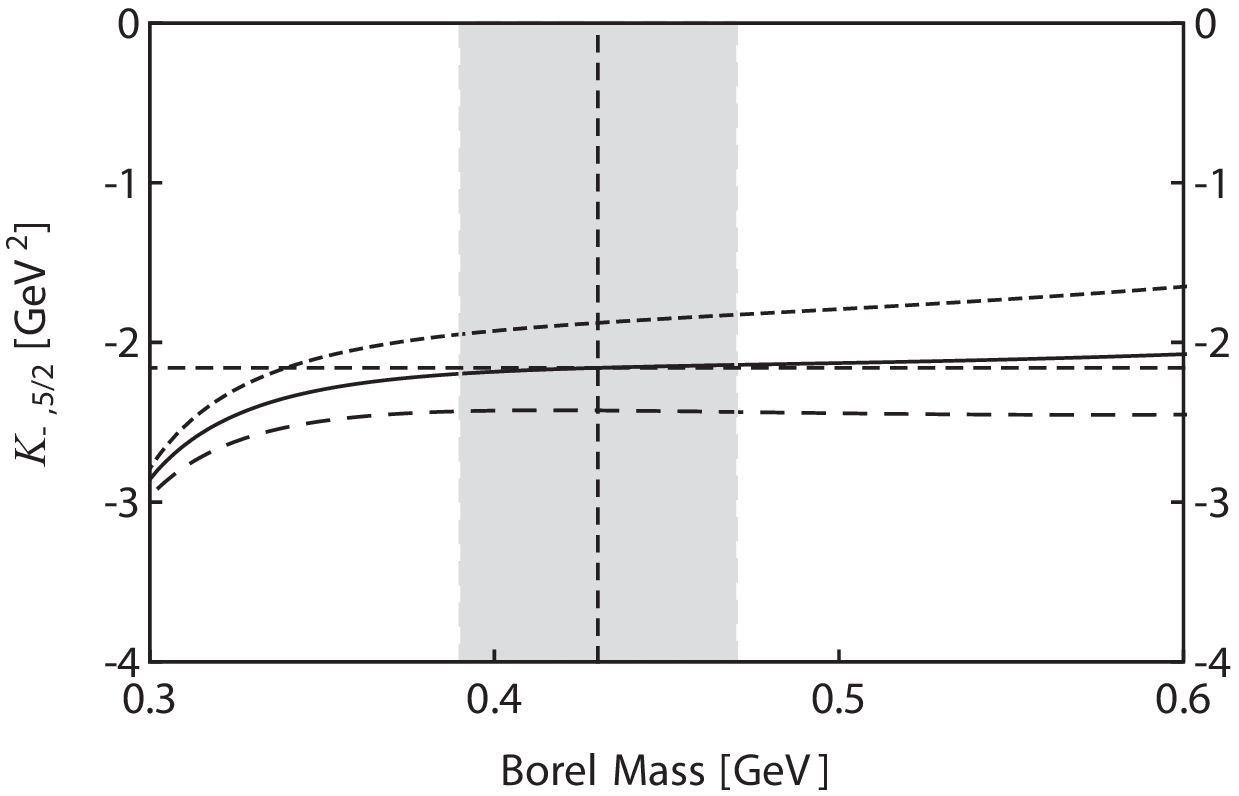}}
\scalebox{0.6}{\includegraphics{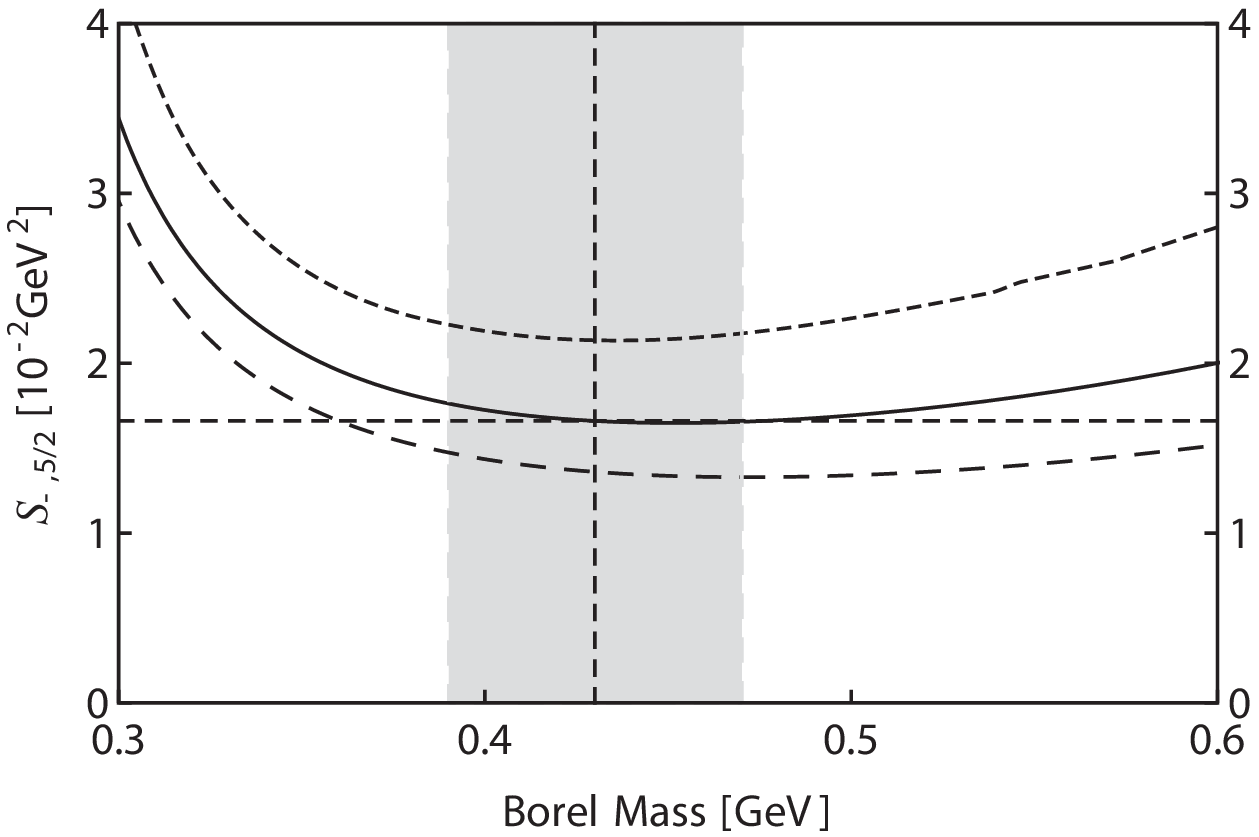}} \caption{The variation
of $K_{-,5/2}$ and $\Sigma_{-,5/2}$ with respect to the Borel mass
$T$ and the threshold value $\omega_c$. The short-dashed, solid and
long-dashed curves are obtained by fixing $\omega_c = 2.5$, 2.7 and
2.9 GeV, respectively. Our working region is $0.39$ GeV $< T < 0.47$
GeV.} \label{fig:next2}
\end{center}
\end{figure}

Similarly we show the variations of $K_{-,5/2}$ and $\Sigma_{-,5/2}$
in Fig.~\ref{fig:next2} in the region $0.30$ GeV $< T < 0.60$ GeV, and their
dependence on the Borel mass $T$ becomes weaker in our working
region $0.39$ GeV $< T < 0.47$ GeV. We obtain the following numerical results:
\begin{eqnarray}
K_{-,5/2} &=& -2.16 \pm 0.28 \mbox{ GeV}^2 \, ,
\\ \Sigma_{-,5/2} &=& 0.017 \pm 0.006 \mbox{ GeV}^{2} \, ,
\end{eqnarray}
where the central value corresponds to $T=0.43$ GeV and $\omega_c =
2.7$ GeV.

\section{Numerical Results and Discussions}
\label{sec:summary}

Combining the results obtained in Sec.~\ref{sec:leading} and
Sec.~\ref{sec:nexttoleading}, we arrive at the following weighted
average mass for the $D$-wave $\bar c s$ heavy meson doublet
$(1^-,2^-)$:
\begin{eqnarray}
{1\over8} ( 3 m_{D^*_{s1}} + 5 m_{D_{s2}} ) &=& m_c + (1.10 \pm
0.06) \mbox{ GeV } \nonumber\\&&+ {1 \over m_c} [ (0.56 \pm 0.09)
\mbox{ GeV}^2 ] \, ,
\end{eqnarray}
where $D_{s2}$ is used to denote the $2^-$ partner of $D_{s1}^*$.
Their mass splitting is:
\begin{eqnarray}
m_{D_{s2}} - m_{D^*_{s1}} &=& {1 \over m_c} [ (0.021 \pm 0.008)
\mbox{ GeV}^2 ] \, .
\end{eqnarray}
From these values, we find that the $\mathcal{O}(1/m_Q)$ corrections
are important and can not be neglected.

To obtain numerical results, we use the PDG value
$m_c = 1.275 \pm 0.025$ GeV~\cite{Agashe:2014kda} for the charm quark mass in the $\overline{\rm MS}$ scheme.
We note that one may also use its pole mass, but then the threshold value $\omega_c$ should be properly
fine-tuned. Therefore, our results for the masses of the heavy mesons have large theoretical uncertainties. However,
their differences within the same doublet do not depend much on the charm quark mass and the threshold value,
so they are produced quite well, with much less theoretical uncertainties:
\begin{eqnarray}
\nonumber m_{D_{s1}^*} &=& 2.81 \pm 0.10 \mbox{ GeV} \, ,
\\ m_{D_{s2}} &=& 2.82 \pm 0.10 \mbox{ GeV} \, ,
\\ \nonumber m_{D_{s2}} - m_{D_{s1}^*} &=& 0.016 \pm 0.007 \mbox{ GeV} \, .
\end{eqnarray}
The mass of $1^-$ state is $2.81 \pm 0.10$ GeV,
consistent with the $D_{s1}^*(2860)$ newly observed by
LHCb, $m^{\rm exp}_{D_{s1}^*} = 2859 \pm 12 \pm 6 \pm 23$
MeV~\cite{Aaij:2014xza}.

Similarly, we obtain the following weighted average mass for the
$D$-wave $\bar c s$ $(2^-,3^-)$ heavy meson doublet:
\begin{eqnarray}
{1\over12} ( 5 m_{D^\prime_{s2}} + 7 m_{D_{s3}^*} ) &=& m_c + (1.14
\pm 0.05) \mbox{ GeV } \nonumber\\&&+ {1 \over m_c} [ (0.54 \pm
0.07) \mbox{ GeV}^2 ] \, ,
\end{eqnarray}
where $D^\prime_{s2}$ is used to denote the $2^-$ partner of
$D_{s3}^*$. Their mass splitting is:
\begin{eqnarray}
m_{D_{s3}^*} - m_{D^\prime_{s2}} &=& {1 \over m_c} [ (0.050 \pm
0.018) \mbox{ GeV}^2 ] \, .
\end{eqnarray}
Again we find that the $\mathcal{O}(1/m_Q)$ corrections are
important and have large uncertainties. Our results are:
\begin{eqnarray}
\nonumber m_{D^\prime_{s2}} &=& 2.81 \pm 0.08 \mbox{ GeV} \, ,
\\ m_{D_{s3}^*} &=& 2.85 \pm 0.08 \mbox{ GeV} \, ,
\\ \nonumber m_{D_{s3}^*} - m_{D^\prime_{s2}} &=& 0.039 \pm 0.014 \mbox{ GeV} \, .
\end{eqnarray}
The mass of the $3^-$ state is $2.85 \pm 0.08$ GeV, also consistent
with the $D_{s3}^*(2860)$ newly observed by LHCb, $m^{\rm
exp}_{D_{s3}^*} =  2860.5 \pm 2.6 \pm 2.5 \pm 6.0$
MeV~\cite{Aaij:2014xza}.

The $\bar bs$ system can be similarly studied by replacing $m_c$ by
$m_b$ and multiplying $\Sigma_{-,j_l}$ by $C_{mag} \approx
0.8$~\cite{Dai:1996qx,Dai:2003yg}. Here we only give their mass
differences within the same doublet because their mass depends much
on the bottom quark mass $m_b$, whose value has large uncertainties.
Using the same threshold values $\omega_c$ around 3.3 GeV and
assuming $4$ GeV$<m_b< 5$ GeV, we obtain the mass differences within
the same doublet
\begin{eqnarray}
m_{B_{s2}} - m_{B_{s1}^*} &=& 0.004 \pm 0.002 \mbox{ GeV} \, ,
\\ \nonumber m_{B_{s3}^*} - m_{B^\prime_{s2}} &=& 0.009 \pm 0.004 \mbox{ GeV} \, .
\end{eqnarray}

We can similarly replace the strange quark by up and down quarks and
extract the masses of the non-strange $D$-wave heavy mesons. To do
this we use slightly smaller threshold values $\omega_c \sim 2.5$
GeV, and obtain the working region $0.39$ GeV $< T < 0.43$ GeV for
$(1^-,2^-)$ doublet. However, there is no stability window for $(2^-,3^-)$
doublet, unless we require the pole contribution to be greater than 20\% only,
and now the working region is $0.46$ GeV $< T < 0.49$ GeV. The numerical results are
\begin{eqnarray}
\nonumber m_{D_{1}^*} &=& 2.75 \pm 0.09 \mbox{ GeV} \, ,
\\ \nonumber m_{D_{2}} &=& 2.78 \pm 0.09 \mbox{ GeV} \, ,
\\ m_{D_{2}} - m_{D_{1}^*} &=& 0.02 \pm 0.01 \mbox{ GeV} \, ,
\\ \nonumber m_{D^\prime_{s2}} &=& 2.72 \pm 0.10 \mbox{ GeV} \, ,
\\ \nonumber m_{D_{s3}^*} &=& 2.78 \pm 0.10 \mbox{ GeV} \, ,
\\ \nonumber m_{D_{s3}^*} - m_{D^\prime_{s2}} &=& 0.06 \pm 0.03 \mbox{ GeV} \, .
\end{eqnarray}
Again we note that the masses have large uncertainties, but their
differences within the same doublet are produced quite well.

In summary, we have studied the $D$-wave  $(1^-,2^-)$ and
$(2^-,3^-)$ $\bar c s$ heavy meson doublets and calculated their
masses up to the $\mathcal{O}(1/m_Q)$ order using the method of QCD
sum rule in the framework of HQET. The masses of $1^-$ and $3^-$
states are calculated to be $m_{D_{s1}^*}$ = $2.81 \pm 0.10$ GeV and
$m_{D_{s3}^*}$ = $2.85 \pm 0.08$ GeV, consistent with the newly
observed $D_{s1}^*(2860)$ and $D_{s3}^*(2860)$ states by
LHCb~\cite{Aaij:2014xza}. In our calculations we have chosen the
same threshold value $\omega \approx 2.7$ GeV for both of them, and
obtained a mass difference between $D_{s1}^*$ and $D_{s3}^*$ to be
0.04 GeV. Considering the mass uncertainties are about 0.1 GeV, our
results are consistent with the experimental
data~\cite{Aaij:2014xza}. The masses of their $2^-$ partners are
calculated to be $2.82 \pm 0.10$ and $2.81 \pm 0.08$ GeV. We note that
our results for the masses of the heavy mesons have large theoretical uncertainties.
One of its sources is the uncertainty of the charm quark mass. Besides this, we do not
consider radiative corrections in the OPE calculations, which may give extra uncertainties.
However, the mass splittings within the same doublet do not depend much on
this, and are reproduced quite well, i.e., $m_{D_{s2}} - m_{D_{s1}^*} = 0.016
\pm 0.007$ GeV and $m_{D_{s3}^*} - m_{D^\prime_{s2}} = 0.039 \pm
0.014$ GeV. We have also estimated their decay constants at the
leading order (in the $m_Q \rightarrow \infty$ limit), that is
$f_{-,3/2} = 0.19 \pm 0.05 \mbox{ GeV}^{7/2}$ and $f_{-,5/2} = 0.15
\pm 0.04 \mbox{ GeV}^{7/2}$.

At present, the two $2^-$ charmed-strange mesons are still missing.
The predicted masses of these two $2^-$ charmed-stange mesons in
this work can be further tested by future experiments. We also
expect more experimental progresses on higher radial and orbital
excitations in the charmed-strange meson family. Besides the experiments, this is also an
ideal system for modern lattice simulations, so we also expect more lattice calculations in this family. We would like to note
that the lattice calculations are becoming more and more precise, and
the precision of their numerical evaluation can be much higher than
in QCD sum rules approach (See Ref.~\cite{Lang:2014yfa} and two very recent talks given in Quarkonium 2014~\cite{mohler:2014,lewis:2014} for more information).

We also obtained the two decay constants $f_{-,3/2}$ and $f_{-,5/2}$, both of which
are important input parameters when performing the dynamical study
relevant to the $D$-wave charmed-strange mesons. With the running of the LHCb experiment, it is an exciting time to
explore the higher charmed-strange mesons. The experimental and
theoretical efforts will establish the charmed-strange meson family
step by step, which is a research area full of challenges and
opportunities.

\section*{Acknowledgments}

This project is supported by the National Natural Science Foundation
of China under Grants No. 11205011, No. 11475015, No. 11375024, No. 11222547, No.
11175073, No. 11035006, and NO. 11261130311, the Ministry of
Education of China (SRFDP under Grant No. 20120211110002 and the
Fundamental Research Funds for the Central Universities), and the
Fok Ying-Tong Education Foundation (No. 131006).


\begin{thebibliography}{99}

\bibitem{Aubert:2003fg}
  B.~Aubert {\it et al.}  [BaBar Collaboration],
  Phys.\ Rev.\ Lett.\  {\bf 90}, 242001 (2003).

\bibitem{Besson:2003cp}
  D.~Besson {\it et al.}  [CLEO Collaboration],
  Phys.\ Rev.\ D {\bf 68}, 032002 (2003)
  [Erratum-ibid.\ D {\bf 75}, 119908 (2007)].

\bibitem{Brodzicka:2007aa}
  J.~Brodzicka {\it et al.}  [Belle Collaboration],
  Phys.\ Rev.\ Lett.\  {\bf 100}, 092001 (2008).

\bibitem{Aubert:2009ah}
  B.~Aubert {\it et al.}  [BaBar Collaboration],
  Phys.\ Rev.\ D {\bf 80} (2009) 092003.

\bibitem{Aubert:2006mh}
  B.~Aubert {\it et al.}  [BaBar Collaboration],
  Phys.\ Rev.\ Lett.\  {\bf 97}, 222001 (2006).

\bibitem{Liu:2010zb}
  X.~Liu,
  Int.\ J.\ Mod.\ Phys.\ Conf.\ Ser.\  {\bf 2}, 147 (2011).

\bibitem{Aaij:2014xza}
  R.~Aaij {\it et al.}  [LHCb Collaboration],
  arXiv:1407.7574 [hep-ex].

\bibitem{Aaij:2014baa}
  R.~Aaij {\it et al.}  [LHCb Collaboration],
  arXiv:1407.7712 [hep-ex].

\bibitem{pdg}K.~A.~Olive {\it et al.} (Particle Data Group), Chin. Phys. C {\bf 38}, 090001 (2014).

\bibitem{Sun:2009tg}
  Z.~F.~Sun and X.~Liu,
  Phys.\ Rev.\ D {\bf 80}, 074037 (2009).

\bibitem{Song:2014mha}
  Q.~T.~Song, D.~Y.~Chen, X.~Liu and T.~Matsuki,
  arXiv:1408.0471 [hep-ph].

\bibitem{Wang:2014jua}
  Z.~G.~Wang,
  arXiv:1408.6465 [hep-ph].

\bibitem{Godfrey:2014fga}
  S.~Godfrey and K.~Moats,
  arXiv:1409.0874 [hep-ph].

\bibitem{Shifman:1978bx}
  M.~A.~Shifman, A.~I.~Vainshtein and V.~I.~Zakharov,
  Nucl.\ Phys.\ B {\bf 147}, 385 (1979).

\bibitem{Reinders:1984sr}
  L.~J.~Reinders, H.~Rubinstein and S.~Yazaki,
  Phys.\ Rept.\  {\bf 127}, 1 (1985).

\bibitem{Grinstein:1990mj}
  B.~Grinstein,
  Nucl.\ Phys.\ B {\bf 339}, 253 (1990).

\bibitem{Eichten:1989zv}
  E.~Eichten and B.~R.~Hill,
  Phys.\ Lett.\ B {\bf 234}, 511 (1990).

\bibitem{Falk:1990yz}
  A.~F.~Falk, H.~Georgi, B.~Grinstein and M.~B.~Wise,
  Nucl.\ Phys.\ B {\bf 343}, 1 (1990).

\bibitem{Bagan:1991sg}
  E.~Bagan, P.~Ball, V.~M.~Braun and H.~G.~Dosch,
  Phys.\ Lett.\ B {\bf 278}, 457 (1992).

\bibitem{Neubert:1991sp}
  M.~Neubert,
  Phys.\ Rev.\ D {\bf 45}, 2451 (1992).

\bibitem{Neubert:1993mb}
  M.~Neubert,
  Phys.\ Rept.\  {\bf 245}, 259 (1994).

\bibitem{Broadhurst:1991fc}
  D.~J.~Broadhurst and A.~G.~Grozin,
  Phys.\ Lett.\ B {\bf 274}, 421 (1992).

\bibitem{Ball:1993xv}
  P.~Ball and V.~M.~Braun,
  Phys.\ Rev.\ D {\bf 49}, 2472 (1994).

\bibitem{Huang:1994zj}
  T.~Huang and C.~W.~Luo,
  Phys.\ Rev.\ D {\bf 50}, 5775 (1994).

\bibitem{Dai:1996yw}
  Y.~B.~Dai, C.~S.~Huang, M.~Q.~Huang and C.~Liu,
  Phys.\ Lett.\ B {\bf 390}, 350 (1997).

\bibitem{Dai:1993kt}
  Y.~B.~Dai, C.~S.~Huang and H.~Y.~Jin,
  Z.\ Phys.\ C {\bf 60}, 527 (1993).

\bibitem{Dai:1996qx}
  Y.~B.~Dai, C.~S.~Huang and M.~Q.~Huang,
  Phys.\ Rev.\ D {\bf 55}, 5719 (1997).

\bibitem{Colangelo:1998ga}
  P.~Colangelo, F.~De Fazio and N.~Paver,
  Phys.\ Rev.\ D {\bf 58}, 116005 (1998).

\bibitem{Dai:2003yg}
  Y.~B.~Dai, C.~S.~Huang, C.~Liu and S.~L.~Zhu,
  Phys.\ Rev.\ D {\bf 68}, 114011 (2003).

\bibitem{Colangelo:1991ug}
  P.~Colangelo, G.~Nardulli, A.~A.~Ovchinnikov and N.~Paver,
  Phys.\ Lett.\ B {\bf 269}, 201 (1991).

\bibitem{Colangelo:1992kc}
  P.~Colangelo, G.~Nardulli and N.~Paver,
  Phys.\ Lett.\ B {\bf 293}, 207 (1992).

\bibitem{feyncalc}
http://www.feyncalc.org/.

\bibitem{Ioffe:2005ym}
  B.~L.~Ioffe,
  Prog.\ Part.\ Nucl.\ Phys.\  {\bf 56}, 232 (2006).

\bibitem{Dai:1998ve}
  Y.~B.~Dai and S.~L.~Zhu,
  Phys.\ Rev.\ D {\bf 58}, 074009 (1998).

\bibitem{Agashe:2014kda}
  K.~A.~Olive {\it et al.}  [Particle Data Group Collaboration],
  Chin.\ Phys.\ C {\bf 38}, 090001 (2014).

\bibitem{Lang:2014yfa}
  C.~B.~Lang, L.~Leskovec, D.~Mohler, S.~Prelovsek and R.~M.~Woloshyn,
  Phys.\ Rev.\ D {\bf 90}, 034510 (2014).

\bibitem{mohler:2014}
D.~Mohler, talk given in Quarikonium 2014, 10-14 November 2014 at CERN,
http://indico.cern.ch/event/278195/session/0/contribution/29/
material/slides/0.pdf.

\bibitem{lewis:2014}
R.~Lewis, talk given in Quarkonium 2014, 10-14 November 2014 at CERN,
http://indico.cern.ch/event/278195/session/0/contribution/30/
material/slides/0.pdf.

\end{thebibliography}
\end{document}